# Object Re-Use & Exchange:
# A Resource-Centric Approach

### Carl Lagoze
Information Science
Cornell University
Ithaca, NY, USA
+1-607-255-6046

lagoze@cs.cornell.edu

### Herbert Van de Sompel
Research Library
Los Alamos National Lab
Los Alamos, NM, USA
+1-505-667-1267

herbertv@lanl.gov

### Michael L. Nelson
Computer Science
Old Dominion University
Norfolk, VA, USA
+1-757-683-6393

mln@cs.odu.edu

### Simeon Warner
Information Science
Cornell University
Ithaca, NY USA
+1-607-254-8605

simeon@cs.cornell.edu

### Robert Sanderson
Department of Computer Science
University of Liverpool
Liverpool, UK
+44 151 795-4252

azaroth@liv.ac.uk

### Pete Johnston
Eduserv Foundation
Bath, UK

+44 1225 474323

pete.johnston@eduserv.org.uk

## ABSTRACT

The OAI Object Reuse and Exchange (OAI-ORE) framework recasts the repository-centric notion of digital object to a bounded aggregation of Web resources. In this manner, digital library content is more integrated with the Web architecture, and thereby more accessible to Web applications and clients. This generalized notion of an aggregation that is independent of repository containment conforms more closely with notions in eScience and eScholarship, where content is distributed across multiple services and databases. We provide a motivation for the OAI-ORE project, review previous interoperability efforts, describe draft ORE specifications and report on promising results from early experimentation that illustrate improved interoperability and reuse of digital objects.

## 1. INTRODUCTION

The Open Archives Initiative (OAI) developed from a Santa Fe meeting in July 1999 to explore interoperability among ePrint archives [41]. According to its original mission inspired by its "roots in the institutional repository movement", OAI "develops and promotes interoperability standards that aim to facilitate the efficient dissemination of content".

The initial work of OAI, the Protocol for Metadata Harvesting (OAI-PMH) [2], reflects this mission and its grounding in mainstream digital library concepts: harvesting *metadata* (primarily bibliographic) from *repositories*. OAI-PMH has been widely deployed, and despite a number of issues related to metadata quality and complexity [23], is considered a successful interoperability mechanism. Its deployment does not compare to related Web-based syndication standards such as RSS and ATOM, due in part to its architectural focus on digital libraries rather than more general Web notions.

In April 2006 Microsoft, the Andrew W. Mellon Foundation, CNI, DLF, and the JISC sponsored a meeting [8] that led to the next and current phase of OAI interoperability work: Object Reuse and Exchange (OAI-ORE). With support of Mellon and Microsoft the work of OAI-ORE has been underway since September 2007, when the following goal was stated: "ORE will develop specifications that allow distributed repositories to exchange information about their constituent digital objects". While this original mission reflects an evolution beyond the metadata-centric nature of OAI-PMH to a focus on content, the mission remains based on core digital library notions, in this case *digital objects stored in repositories* [20].

This paper describes the results of one-and-a-half years of OAI-ORE work, a set of specifications and user guides [26] that state: "Open Archives Initiative Object Reuse and Exchange (OAI-ORE) defines standards for the description and exchange of aggregations of Web resources." This represents yet another evolution of the OAI mission: from a *repository-centric* focus and a conceptualization of content as *stored* in repositories, which has characterized most digital library work, to a *resource-centric* focus in which machines (e.g. Web servers) act as service points to content independent of location. The salient aspects of the conceptual differences between OAI-PMH to OAI-ORE are illustrated in Table 1.

**Table 1 - Concepts in OAI-PMH vs OAI-ORE**

| OAI-PMH | OAI-ORE |
|---|---|
| Repository structure | Object structure |
| Repository-centric | Web-centric |
| Metadata-centric | Resource-centric |
| Metadata harvesting | Object re-use (using URI as "handle") |

This evolution of goals reflects the participation in the development of ORE of experts from a variety of communities beyond digital libraries, including Web Architecture, eScience, Semantic Web, and others. It also reflects a recognition of the position of digital libraries vis-à-vis the Web that sometimes seem to co-exist in a curiously parallel conceptual and architectural space.

We exist in a world where information is synonymous not with "library" but with the Web and the applications that are rooted in it. In this world, the *Web Architecture* is the lingua franca for information interoperability, and applications such as most digital libraries must exist within the capabilities and constraints of that Web Architecture. Because of the virtual hegemony of Web browsers as an information access tool and Google as a discovery tool, failure to heed to Web Architecture principles, and therefore requiring somewhat special treatment by these "monopoly applications" (which is rarely if ever granted), effectively means falling into an information black hole.

Full details on the Web Architecture are described in [19]. Stated briefly, it provides the following notions:

- *Resource* - an item of interest.
- *URI* - a uniform global identifier for a Resource. URIs comply to URI schemes (e.g., http, ftp, gopher) and each scheme defines the mechanism for assigning URIs within that scheme. Within the common http scheme, the URI is an identifier key in an HTTP (hypertext transfer protocol) request message, which may result in the return of information about the respective Resource. However, the ability to automatically *de-reference* an http URI is not true for all URIs (nor even for all http URIs).
- *Representation* - a data stream corresponding to the state of a Resource at the time its URI is dereferenced. The Web Architecture allows for multiple Representations of a Resource with access mediated by *Content Negotiation*.
- *Link* - a directed connection between two Resources. In most common usage, a link is expressed via link or anchor tags (a *hyperlink*) in an HTML Representation of the originating Resource to the URI of another Resource. An extension of this, where links are typed relationships, is one of the goals of the *Semantic Web*.

The notion of a *web server*, which is somewhat analogous to the digital library notion of a *repository*, is not included in the Web Architecture. This does not mean that the digital library notion of a repository is irrelevant, and in fact we argue that issues essential to digital libraries such as preservation, authority, and integrity largely rely on the repository as a management entity. However, a repository-centric approach to interoperability may produce results that do not coordinate well with the resource-centric architecture of the Web, leading to the "black hole" scenario mentioned above.

The digital library notion of a *digital object*, in particular one that is a compound *aggregation,* is another concept without strict equivalence in the Web Architecture. The repository technologies that originally motivated the ORE work, such as DSpace, Fedora, aDORe, ePrints and arXiv, all store content that is more than a simple file, albeit, they differ in how they implement this and in the richness of their functionality. A look at the arXiv for example shows that most content is available in multiple formats (e.g., PDF, LaTeX), is versioned, is represented by some metadata format, and has citations to other papers. Collectively this aggregation of elements is the "document" in arXiv.

While the notion of an *aggregation* is not explicit in the Web Architecture, it is prevalent across general Web space. For example, a "photo" in Flickr is an aggregation of multiple renditions in different sizes, and that photo is aggregated along with other "photos" into a "collection". Similarly, the blog entry that we think of as a singleton is in fact an aggregation composed of the original entry combined with multiple comments (and comments on comments). That blog entry is itself aggregated in a subject partition of a blog.

Turning to the eScience/eScholarship context, which is increasingly the focus of digital library activities, we see more examples of *aggregations*, with components that are distributed across multiple services and databases. For example the multi-part "virtual data" objects envisioned by the National Virtual Observatory Project [43], the "datuments" described in the chemistry community [30] and the learning objects implemented by NSDL [24] all share the property that their components are distributed over multiple databases, web servers, databases, and the like. In this context, the notion of a repository as a container is not especially relevant. Rather content is distributed and made available via distributed service points.

While aggregations are prevalent in the Web, their absence from the architecture strips them of two properties fundamental to digital objects in digital libraries:

- *Identity:* Digital objects have identifiers such as handles or DOIs that identify the *whole object.* This identity is important as the means of expressing citation, lineage, and rights. We argue that it is also relevant in the Web context, especially in the Semantic Web where identities are the subjects and objects of RDF assertion, and an assertion about a splash page needs to be distinct from an assertion about an aggregation as a unit.
- *Boundary:* A fundamental aspect of a digital object is that it is possible to deterministically enumerate its constituents. This is vital for services such as preservation (what to preserve) and rights management (who is responsible for what). While not defined in the Web Architecture, the importance of boundary has also been acknowledged in Web applications. It is therefore part of the requirement set of the Protocol for Web Description Resources (POWDER) [4] work, which aims to provide mechanisms to publish properties shared by a set of Web resources.

The ORE work is motivated by this generality of aggregations, with identity and boundary, across all web-based information systems, including digital libraries and eScience/eScholarship applications. At the time of writing this paper, the ORE specifications are still in alpha status and, while they have been the subject of a number of experiments (described later in this paper), real applications that exploit them have yet to be built. However, we propose the following applications for the machine-readable descriptions of aggregations defined by OAI-ORE:

- Crawler-based search engines could use such descriptions to index information and provide search results sets at the granularity of the aggregations rather or in addition to their individual parts.
- Browsers could leverage them to provide users with navigation aids for the aggregated resources, in the same manner that machine-readable site maps provide navigation clues for crawlers.
- Other automated agents such as preservation systems could use these descriptions as guides to understand a "whole document" and determine the best preservation strategy.
- Systems that mine and analyze networked information for citation analysis and bibliometrics could achieve better accuracy with the knowledge of aggregation structure contained in these descriptions.
- These machine-readable descriptions could provide the foundation for advanced scholarly communication systems that allow the flexible reuse and refactoring of rich scholarly artifacts and their components [40].

The next section describes related interoperability work on compound digital objects, object description formats, repository architecture, interoperability protocols, and identification. A section that describes the ORE specifications follows. Next, we describe ongoing experiments with the specifications. Finally, we address some conclusions and challenges for the future of our work.

## 2. RELATED INTEROPERABILITY WORK

OAI-ORE is best understood within the context of prior and continuing web-based DL interoperability efforts. This section will give a brief overview of the projects, formats and protocols that have shaped the design of ORE.

Any discussion of DL interoperability is likely to begin with what is informally known as the Kahn-Wilensky Framework (KWF) [20]. Originally published as a web page in 1995, the KWF was the architecture for the Computer Science Technical Report (CS-TR) project [5]. The CS-TR project later merged with the WATERS project [28] to form the basis for the Dienst protocol [22] and the NCSTRL project [16]. Lessons learned with Dienst and NCSTRL later significantly influenced the design of OAI-PMH.

The KWF influenced some of the thinking in the Dublin Core (DC) community, resulting in the Warwick Framework [21], which was later extended with "distributed active relationships" [15], which later evolved into Fedora [25]. The KWF also formed the basis for a prototype implementation for the Library of Congress National Digital Library Program [6]. The representation of metadata in digital objects in the NDLP influenced the Making of America II project [17], which gave rise to the Metadata Encoding and Transmission Standard (METS) [29].

The KWF also popularized an identifier scheme known as handles [37]. The handle system is widely used and is the key technology in the implementation of digital object identifiers (DOIs) [32].

As the above history indicates the influence of the KWF has been extensive and its contributions can be grouped into the areas of 1) repository protocols, 2) digital objects and 3) identifiers. In the subsections below we explore each of these topics further, starting with their origins and continuing to their present status and influence on ORE.

### 2.1 Repository Protocols

Perhaps owing to the influence of pre-Web Z39.50, early DL protocols approached interoperability via support of distributed (or "federated") searching. The aforementioned Dienst protocol provided many things, including: mediated access to holdings in a repository conformant to a structured data model, bibliographic metadata exchange and support for distributed searching. While Dienst provided interoperability with other Dienst implementations, other projects such as the Stanford Simple Digital Library Interoperability Protocol [18], attempted to provide interoperability between heterogeneous systems (e.g. Dienst, Z39.50, etc.) by providing a generic, "wrapper" protocol that abstracted the shared semantics between various systems. A similar project, Stanford Protocol Proposal for Internet Retrieval and Search (STARTS) [18], defined a method for repositories to expose just enough information about their holdings and capabilities to facilitate distributed searching.

After several years it became apparent that for both theoretical and engineering reasons, achieving repository interoperability through large-scale distributed searching was difficult (e.g., [34]). The OAI-PMH was informed by these experiences and its approach to interoperability was intentionally limited to bibliographic metadata exchange, effectively based on a small subset of the larger Dienst protocol [22]. Although (as mentioned above) the OAI-PMH data model contains constructs that do not directly map to the Web Architecture, the OAI-PMH approach to interoperability has more in common with current Web approaches such as syndication formats (e.g., RSS and Atom) and SiteMaps than it does with previous efforts at repository interoperability. It reflects a simple approach to repository interoperability dependent on enumerating and describing holdings (it is up to the consuming applications to build services on the exposed resources).

### 2.2 Digital Objects

In response to this simplicity, complexity and expressiveness has moved from the protocols to the formats of the digital objects. The concept of digital objects, including typed, recursive and composite digital objects, is fundamental to the KWF. Drawing from Arm's observation that "users want intellectual works, not digital objects" [5], repositories have co-developed with object description formats to describe and manage these "intellectual works" (or "works" and "expressions" in FRBR terminology [1]).

As mentioned above, METS has a lineage back to the KWF, and is (or was) the default object description format for many repository projects, such as DSpace [36] and Fedora. Other communities have created or adopted their own object formats: IMS-LOM [33], from the Learning Objects community, and MPEG-21 DIDL, originally from the consumer electronics community and adapted to the DL environment by Los Alamos National Laboratory [7]. Although the syntax and application domain for these formats differ, they all have goal of combining descriptive, structural and administrative metadata to conjure digital objects of "intellectual works".

Although OAI-PMH has its origins in the harvesting of descriptive metadata, OAI-PMH has been combined with object formats such as METS and DIDL to create "resource harvesting" [42]. This has been studied in the context of transferring digital objects between repositories in the APS-LANL project, effectively combining OAI-PMH and Open Archival Information System (OAIS) reference model [9].

Despite their utility, the current widespread use of these object description formats presents at least two problems. First, as the expressiveness and complexity of a format increases, interoperability becomes more difficult. For example, in the Archive Ingest and Handling Test [35] the four participants ingested the same resources in their respective, differing repositories. When they encoded their contents for export (3 in METS, 1 in MPEG-21 DIDL), none of the parties could ingest the export of the others without significant pre-processing; format expressiveness had come at the cost of at least initial interoperability. Secondly, there is no clear mapping of these compound objects into the Web Architecture. To borrow from FRBR terminology again, object description formats, and the identifiers they use, are primarily about "works" or "expressions" and the Web Architecture is primarily about manifestations (resources) and items (representations).

## 2.3 Identifiers

While "names and identifiers are the basic building block for the digital library" [5], there are a number of fundamental tensions present in the use of identifiers. DOIs have been a significant catalyst for interoperability in the scholarly publishing community. But their ubiquity underlies their ambiguity: in the context of the Web, what do they actually identify? This is really the larger question of resolvable and non-resolvable identifiers. From the DL perspective, there is significant value in the ability of a non-resolvable identifier such as `info:doi/10.1007/s00799-007-0016-7` to identify an intellectual work, but from a Web perspective this identifier is of limited use without employing gateways such as `http://dx.doi.org` or OpenURL resolvers for service. It is common to use URIs such as `http://dx.doi.org/10.1007/s00799-007-0016-7` to identify both the intellectual work and the HTML "splash page" returned when the URI is resolved (a FRBR "item").

Although this ambiguity is not a problem in conventional browsing (humans can often distinguish when the URI is identifying the intellectual work and when it is identifying an HTML page), it does hinder the development of automated services that do not always understand the subtle convention that `http://arxiv.org/abs/cs/0610031v1` is in fact just one of many members of the intellectual work properly identified by `info:arxiv:cs/0610031v1` and not the intellectual work itself. The present ambiguity of allowing, depending on context, the former URI to represent both a set and a member of a set is one of the remaining fundamental problems of interoperability.

## 3. ORE SPECIFICATIONS

The ORE specifications aim to address the interoperability issues raised in the related work described in the previous section. The ORE alpha specifications were made public on 10 December 2007 [26] for a period of review and consultation. Discussion groups, meetings and experimentation will guide evolution through beta to final specifications, the release of which are expected in 3rd quarter 2008. The suite of documents contains both specifications and user guide documents. We focus here on three key aspects: the data model, serialization, and discovery.

The object of the ORE specifications is to add *machine-readable* information to the Web that augments the human-readable Web. Various discovery mechanisms provide hooks whereby browsers and agents surfing the human-readable Web can find out about ORE information which may then be used to direct or augment the functions available (e.g. "print whole chapter" from a web page displaying a page image). The central notion of an aggregation adds *boundary* information to a set of web resources that may be arbitrarily distributed over many servers (e.g. a large dataset, model code, an article, and open-review commentaries).

## 3.1 ORE DATA MODEL

The ORE Data Model makes it possible to associate an identity with aggregations of Web resources and to describe their structure and semantics. It does this by introducing the *Resource Map* (ReM), which is a resource identified by a URI (say `ReM-1`) that encapsulates a set of RDF statements[1]. The notion of associating a URI with a set of RDF statements is based on the concept of a *named graph* developed in the Semantic Web community [12]. The creation of a Resource Map instantiates an aggregation as a resource with a URI distinct from the Resource Map, enumerates the constituents of the aggregation, and defines the relationships among those constituents.

It is important to note that an Aggregation described by a Resource Map is independent of other notions of aggregations or compound digital objects in repositories or other servers. An ORE Aggregation exists only in tandem with, and in fact, due to the existence of a single Resource Map. As described below, this binding is enforced by the URI syntax of Resource Maps and Aggregations. Also, the sections below describe the means of establishing linkages between an Aggregation and digital objects in other architectural contexts.

RDF triples in the ORE Model use predicates from a number of vocabularies and a few additions from the ORE namespace. Table 2 shows the namespace prefixes used in the ORE specifications and in this paper.

**Table 2 - ORE namespace prefixes**

| Prefix | Namespace URI | Description |
|--------|---------------|-------------|
| dc | `http://purl.org/dc/elements/1.1/` | Dublin Core elements |
| dcterms | `http://purl.org/dc/terms/` | Dublin Core terms |
| ore | `http://www.openarchives.org/ore/terms/` | ORE vocabulary terms |
| owl | `http://www.w3.org/2002/07/owl#` | OWL vocabulary terms |
| rdf | http://www.w3.org/1999/02/22-rdf-syntax-ns# | RDF vocabulary terms |

### 3.1.1 Aggregation

A Resource Map describes an *Aggregation* which is a set of resources; the Resource Map may provide information about the types of the resources and relationships among them. Resources in the Aggregation are called *Aggregated Resources*.

In order to be able to reference the Aggregation on the Web, it must have a URI (say `A-1`). This URI is constructed by appending "#aggregation" to the Resource Map URI, i.e. `ReM-1#aggregation`. This syntactic device ensures that there is a unique Aggregation resource for every Resource Map. Moreover, it makes it possible to always find the Resource Map given the URI of the Aggregation. The "#aggregation" naming convention is also consistent with how the Semantic Web community assign URIs for "non-informational resources" (i.e., resources that exist outside of the Web: people, places, concepts – including such abstract things as FRBR works) [11].

The relationship between URIs of common "splash pages" and URIs for digital objects or aggregations to which these splash pages are entry points is often ambiguously defined in the repository community. The ORE specifications emphasize that these concepts should not be conflated and that they should have separate URIs. This separation is the only manner in which assertions about them can remain distinct. However, it is likely and appropriate that many repository systems will include splash pages as an Aggregated Resource in an Aggregation, but they should *not* consider a splash page as one representation of the Aggregation.

---

[1] RDF statements are triples that relate a subject resource to an object resource or literal via a predicate (relationship). In this section we illustrate ORE data model concepts using the simple N3 format [10] Berners-Lee, T. Notation 3, World Wide Web Consortium, 1998. of "subject object predicate." for each triple.

### 3.1.2 Resource Map

Figure 1 shows a simple Resource Map for a version of a paper on arXiv. In the figure, individual RDF statements consisting of a subject resource and object resource or literal connected via an arrow that is the predicate are joined in a single graph. The remainder of this section builds and explains the components of this graph step-by-step. The key components – Resource Map, Aggregation and Aggregated Resources – are shown in black and we use the shorthand labels in place of the full URIs.

The Resource Map is identified by `ReM-1` and an HTTP GET on `ReM-1` must yield a serialization of the Resource Map. Note also that ReM-1 appears as a node in the graph and is the subject of several triples. First, there must be triples stating that resource `ReM-1` is a Resource Map, that resource `A-1` is an Aggregation, and linking the Resource Map to the Aggregation that it describes:

```
<ReM-1>   rdf:type       ore:ResourceMap.
<A-1>     rdf:type       ore:Aggregation.
<ReM-1>   ore:describes  <A-1>.
```

There are two mandatory metadata elements about the Resource Map (not the Aggregation), the authority or person that created the Resource Map and the last modified date:

```
<ReM-1>   dc:creator        <http://arxiv.org/>.
<ReM-1>   dcterms:modified
          "2008-01-15T10:01:19Z".
```

Two additional metadata elements about the Resource Map may optionally be included:

```
<ReM-1>   dc:rights
          <http://creativecommons.org/publicdomain/>.
<ReM-1>   dcterms:created "2008-01-15T10:01:19Z".
```

The two Aggregated Resources are linked to the Aggregation with the `ore:aggregates` predicate and may optionally be typed as `ore:AggregatedResource`:

```
<A-1>     ore:aggregates  <AR-1>.
<AR-1>    rdf:type        ore:AggregatedResource.
<A-1>     ore:aggregates  <AR-2>.
<AR-2>    rdf:type        ore:AggregatedResource.
```

Thus far, the Aggregation is just a bag of resources, `AR-1`, and `AR-2`, unrelated and not described except for their status as constituents of the Aggregation, A-1. There are significant applications where this is already useful: for example the notion of grouping in intellectual objects used by Google Scholar -- links to the splash page, PDF and HTML version of an article should be considered links to the same intellectual object. However, in many cases additional description will be useful.

If the Aggregation denotes or is similar to a resource that has other identifiers, then these are expressed using either the `owl:sameAs` or `ore:analogousTo` predicate. It is important to understand that `owl:sameAs` makes a strong statement of equivalence between two URIs: they identify the same resource and thus one URI may be substituted for the other. We introduce the weaker relation, `ore:analogousTo`, which implies equivalence of two resources without substitutability. In the example, Aggregation `A-1` represents the same intellectual object as a version published in a journal with a DOI. This and the semantic type are indicated with:

```
<A-1>     ore:analogousTo
          <info:doi/10.1103/PhysRevD.72.095016>.
<A-1>     rdf:type
          <http://purl.org/eprint/type/JournalArticle>.
```

A Resource Map may also describe the structure of the Aggregation by expressing relationships between the

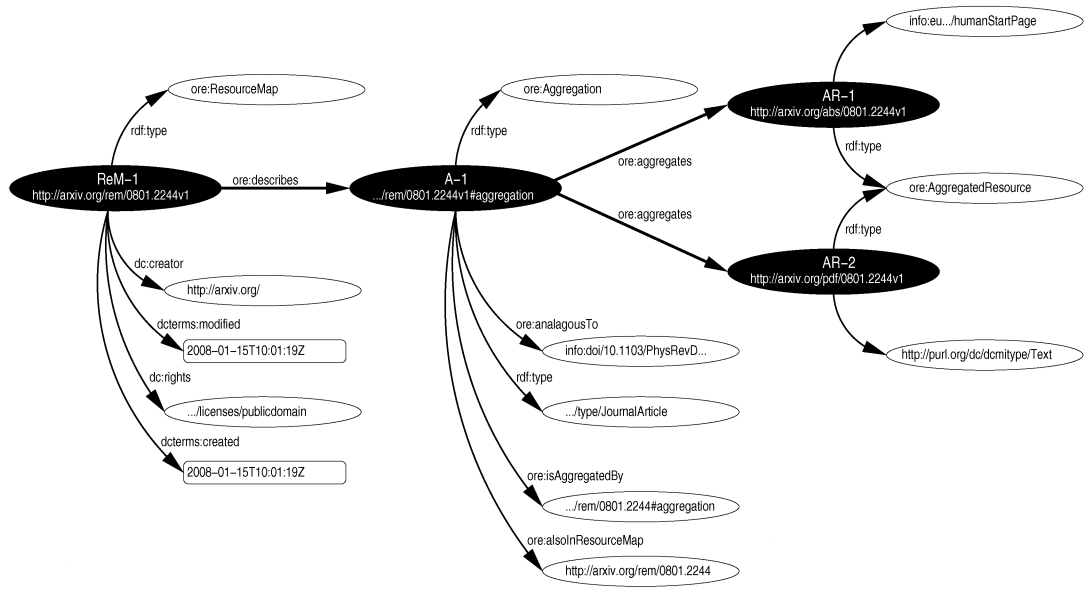

**Figure 1 – An Example OAI-ORE Resource Map**

Aggregation, the Aggregated Resources and other resources. For brevity, the example includes just type information of the two Aggregated Resources:

```
<AR-1> rdf:type
    <info:eu-repo/semantics/humanStartPage>.
<AR-2> rdf:type
    <http://purl.org/dc/dcmitype/Text>.
```

Additional statements might relate components to each other (e.g. `dcterms:isPartOf` to indicate part-whole) and to other resources (e.g. `dcterms:references` to indicate citation). A number of recommended vocabulary terms are included in the ORE specification to promote interoperability, but the use of terms from any vocabulary is permitted.

### 3.1.3 Relationships to other Aggregations

When reusing Resource Maps and the Aggregations that they describe, it is important to remember the distinction between the two concepts. Statements about `ReM-1` are statements about the Resource Map and not the Aggregation; statements about `A-1` (= `ReM-1#aggregation`) are statements about the intellectual object that is the Aggregation.

One common relationship between aggregations is nesting. In the arXiv example, the Resource Map describes a particular version of an article and the graph shows that this aggregation is nested within another aggregation representing all versions of the article using:

```
<A-1>  ore:isAggregatedBy
    <http://arxiv.org/rem/0801.2244#aggregation>.
<A-1>  ore:alsoInResourceMap
    <http://arxiv.org/rem/0801.2244>.
```

In the more general case, any resource may be aggregated in more than one Aggregation, each described by a Resource Map (say `ReM-1` and `ReM-2`). To support discovery, the predicate `ore:alsoInResourceMap` allows specifying that an Aggregated Resource from one Resource Map is also an Aggregated Resource in another Resource Map. For example, `ReM-1` might contain the following triple expressing that `AR-1` is known to also be aggregated in `ReM-2` (not shown in figure):

```
<AR-1>      ore:alsoInResourceMap      <ReM-2>.
```

At the time of writing, a second relationship to other Resource Maps is still the subject of discussion. It is the lineage relationship to indicate that an Aggregated Resource originates in another aggregation. In this case the predicate is `ore:fromResourceMap`, for example (not shown in figure):

```
<AR-1>      ore:fromResourceMap      <ReM-3>.
```

A problem with the proposed predicate `ore:fromResourceMap` is that it should only be interpreted in the context of the asserting Resource Map. Standard RDF models (triples) don't support this notion but systems that retain context information (quad stores etc.) can. Systems than cannot understand context should interpret `ore:fromResourceMap` in the same way as `ore:alsoInResourceMap` which is less expressive but correct.

## 3.2  Serialization

In order to bootstrap the deployment of ORE Resource Maps, a serialization based on the Atom Syndication Format [31] is explicitly specified. The choice to promote a single format was motivated by the desire to promote interoperability without the need for format conversions. Atom was chosen because it matches well the concepts of the ORE model and is a widely used-format that is gaining adoption in a number of areas beyond the typical blog applications (e.g. Google Data[2]).

### 3.2.1 Atom serialization

An Atom feed aggregates a number of entries. When using Atom for ORE, a Resource Map is mapped to an Atom feed, and each Aggregated Resource to an Atom entry. The four metadata elements about the Resource Map are provided using feed-level Atom metadata elements. The rules for mapping all entities of the ORE Model to and from Atom are described in detail in the specification. Here we illustrate the key points with the example shown in Figure 2 which is a Resource Map for an arXiv e-print with just two components shown: a PDF version and a HTML splash page.

At the feed level, the Resource Map and Aggregation URIs are indicated with `/feed/link[@rel="self"]` and `/feed/link[@rel="describes"]` elements, respectively. The relation "`self`" follows standard Atom semantics but "`describes`" is an ORE addition[3] to indicate the Aggregation described by the feed. The mandatory modification time and creator metadata elements map to the Atom `/feed/updated` and `/feed/author` elements, respectively. The `/feed/author` element admits `name`, `uri` and `email` sub-elements. Only the `name` or `uri` sub-elements have meaning in the ORE model and are mapped to the `dc:creator` triple with either a literal (`name`) or a resource (`uri`) as the object of the triple.

The URI of the Aggregated Resource is indicated with `/feed/entry/link[@rel="alternate"]`. Additional triples associated with an Aggregated Resource are included as child elements of this `/feed/entry` using elements from non-Atom namespaces. The example shows `rdf:type` elements for `AR-1` and `AR-2`.

Atom mandates that feeds and entries have globally unique URIs (`/feed/id` and `/feed/entry/id`) and some additional metadata (e.g. `/feed/title` and `/feed/entry/title`); these have no correspondence in the ORE model. Feed creating applications must mint these URIs to produce valid Atom feeds and should be careful that they are globally unique and persistent, but must not reuse the Aggregation and Aggregated Resource URIs. For the feed and entry titles it is recommended to use the Resource Map and Aggregated Resource URIs, prefixed with "`Resource Map`" and "`Aggregated Resource`" to provide a human readable description of the content.

The Atom format is extensible and we make use of this feature in serializing core elements of the ORE Data model as described above. Arbitrary elements from other namespaces, including RDF, are permitted within Atom feed documents so it is possible to create an Atom serialization that expresses relationships among aggregated resources. However, because these are extensions without standard ATOM semantics, conventional Atom applications will effectively ignore them.

---

### 3.2.2 Other serializations and round trip behavior

The choice of Atom as the primary serialization format is not intended to preclude the use of other serializations. However, different serializations may be able to represent aggregations conforming to the ORE data model with differing degrees of fidelity. Clearly, any format capable of serializing an arbitrary RDF graph can be used to serialize a Resource Map with complete fidelity, and examples include N3, RDF/XML, Trix, and Trig. As mentioned above, Atom serialization for Resource Maps is less expressive, and can, for example, not express a relationship where an Aggregated Resource is the object (instead of subject) of a relationship triple.

For any serialization to be useful there must be a well-defined bi-directional mapping to the ORE Model. A test of this mapping is that one must be able to make the round trip between the model and representation without data loss or corruption. However, because of the possibility of both limited expressiveness and/or of additional features in a particular serialization we must be careful to define the round trip. The mapping must preserve intact all information on the *second* and subsequent round trips. For example, to check the mapping to format X one must find the common expressiveness by doing the first round trip model→X→model (or X→model→X), and then verify that an additional round trip model→X→model (or X→model→X) preserves

```
<?xml version="1.0" encoding="utf-8"?>
<feed xmlns="http://www.w3.org/2005/Atom">
  <id>tag:arxiv.org,2007:astro-ph/0601007v2</id>
  <link href="http://arxiv.org/rem/astro-ph/0601007v2"
    rel="self" type="application/atom+xml"/>
  <category scheme="http://www.openarchives.org/ore/terms/"
      term="http://www.openarchives.org/ore/terms/ResourceMap"
      label="Resource Map"/>
  <link rel="describes"
      href="http://arxiv.org/rem/astro-ph/0601007v2#aggregation"/>
  <title>Resource Map http://arxiv.org/rem/astro-ph/0601007v2</title>
  <author>
    <name>arXiv.org e-Print Repository</name>
    <uri>http://arxiv.org/</uri>
    <email>www-admin@arxiv.org</email>
  </author>
  <updated>2007-10-10T18:30:02Z</updated>
  <entry>
    <id>tag:arxiv.org,2007:astro-ph/0601007v2:ps</id>
    <link href="http://arxiv.org/ps/astro-ph/0601007v2"
      rel="alternate"
      type="application/postscript"/>
    <title>Aggregated Resource http://arxiv.org/ps/astro-ph/0601007v2</title>
    <updated>2006-05-31T12:52:00Z</updated>
  </entry>
  <entry>
    <id>tag:arxiv.org,2007:astro-ph/0601007v2:pdf</id>
    <link href="http://arxiv.org/pdf/astro-ph/0601007v2"
      rel="alternate" type="application/pdf"/>
    <title>Aggregated Resource http://arxiv.org/pdf/astro-ph/0601007v2</title>
    <updated>2006-05-31T12:52:00Z</updated>
  </entry>
  ...
</feed>
```

**Figure 2 – Resource Map in Atom**

all information.

## 3.3 Discovery

Discovery is a precondition for the use of Resource Maps. There is no single, best method for discovering Resource Maps, and we expect best practices for discovery to evolve over time. The Resource Map Discovery Document [27] covers a variety of suggested Resource Map discovery mechanisms, grouped into the categories of Batch Discovery, Resource Embedding and Response Embedding.

### 3.3.1 Batch Discovery

Batch discovery exists so agents can discover Resource Maps *en masse*. Note that Resource Maps are not limited to describing Aggregations on the server where the Resource Maps reside. This means that a machine in domain A can make Resource Maps available that describe aggregations of resources from domains B, C and D. Assuming the Aggregated Resources are not remotely editable, batch discovery techniques are the most direct method of publishing third party aggregations.

It is possible to discover Resource Maps from an OAI-PMH repository. We anticipate this will allow rapid ORE adoption in the repository community. The most likely method would be to define a new metadataPrefix (`oai_rem`) that could be used to harvest Resource Maps associated with corresponding OAI items. Similarly, Resource Maps could be made available in SiteMaps, either in a separate SiteMap exclusively listing Resource Maps or a SiteMap with regular resources and Resource Maps interspersed. Even though the preliminary serialization of Resource Maps is the Atom Syndication Format, there is no reason preventing the use of syndication formats such as Atom or RSS for Resource Map discovery. However, care must be taken to separate conceptually the Resource Map from the syndication file listing the Resource Maps. For example, the id of an Atom entry listing the URI of a Resource Map must be neither the URI of that Resource Map nor the Atom feed id of the Resource Map. Furthermore, an explicit difference must be made between the Atom feed used for discovery and the Atom feed that is the Resource Map. Further guidance for syndicating Resource Maps is provided in the Discovery Document [27].

### 3.3.2 Resource and Response Embedding

Individual resources can inform ORE-aware applications about (at least some of) their corresponding Resource Maps. HTML resources can use the `<link>` element to point to Resource Maps that describe the Aggregations of which they are an Aggregated Resource. Non-HTML resources can use the proposed HTTP `<link>` response header, which is functionally equivalent to the `<link>` element. These techniques are referred to as resource and response embedding of Resource Maps, respectively.

The Resource Map link (via either the HTML element or HTTP response header) can be used to direct agents from the Aggregated Resource to a corresponding Resource Map that describes the Aggregation of which the resource is part. While this is a common case, there are actually four different scenarios regarding members of an Aggregation and knowledge about their corresponding Resource Maps:

- Full knowledge: all resources in the Aggregation link to the Resource Map.

- Indirect knowledge: all but one of the resources in the Aggregation link to a single, unique resource in the Aggregation, which in turn links to the Resource Map.
- Limited knowledge: only a subset of the resources in the Aggregation (typically just a single resource) link to the Resource Map, and the remainder of the resources have no links at all.
- Zero knowledge: none of the resources in the Aggregation link to a Resource Map.

Note that the above scenarios are relative to a particular Resource Map. It is possible for Aggregated Resources to simultaneously have full knowledge about one Resource Map (typically authored by the same creators of the resources) and have zero knowledge about third party Resource Maps that describe aggregations of the same resources. Full, indirect or limited knowledge can be interpreted as the Resource Map being "endorsed" by the resource creator. However, there is no concept of a "negative endorsement" — zero knowledge could mean the creators either do not endorse the Resource Map or are simply unaware of the Resource Map.

# 4. EXPERIMENTATION AND IMPLEMENTATION

Early experimentation is an essential tool to assess the real-world feasibility of evolving specifications and of the architectural paradigms on which they are based. Therefore, several members of the ORE Technical Committee have engaged in small-scale explorative projects.

In January 2007, three months into the ORE effort, the Digital Library Research & Prototyping Team of the Los Alamos National Laboratory (LANL) conducted an experiment in which the Zotero citation manager browser plug-in [13] was modified to detect the existence of a compound information object from the HTML splash page for a scholarly article. When detected, the enhanced Zotero offered the user the ability to download any number of constituent resources of the compound object, including, obviously, its bibliographic description. In this experiment, compound information objects were represented as special-purpose ATOM feeds. Leveraging ATOM as a strategy to integrate compound scholarly objects into the mainstream Web has remained a theme throughout the ORE effort.

In May 2007, before the outlines of the ORE data model to handle aggregations of Web resources were stabilized, the LANL Team conducted the Compound Information Object Archive Prototype experiment. A movie describing this experiment is available at [39]. The experiment demonstrates archiving compound information objects as they evolve over time. Each object is published at a stable http URI from which an ATOM-based Resource Map is available. A Resource Map lists the http URIs of the constituents of the compound object at the time of retrieving its ATOM representation. A local implementation of the Internet Archive recurrently collects both the Resource Maps and their listed constituents. In essence, the experiment shows that, without any modification, an existing web application — the Internet Archive — can take advantage of published Resource Maps to effectively create a self-contained archive of evolving compound objects. The core enabler is the publication of Resource Maps that describe compound objects in a machine-readable way at stable http URIs. This has become a central concept in the ORE data model.

In the Australia-based SCOPE project, Jane Hunter and colleagues started leveraging ideas that had crystallized during the May 2007 meeting of the ORE Technical Committee. SCOPE aims at providing researchers with a simple desktop tool to construct scholarly compound objects consisting of any number of Web resources (e.g. objects from repositories). But, to meet eScience requirements, SCOPE also focuses on recording provenance information for those resources, as well as on maintaining details regarding the workflow that was used to generate one constituent resource on the basis of another. In an experiment reported in [14], SCOPE uses the Named Graph concept that has become central to ORE to model these compound objects, and serializes these Named Graphs, among others, in RDF/XML and ATOM. These serializations can then be submitted to a repository, where the compound object represented by the serialization can be maintained.

In November 2007, the LANL Team conducted another experiment, this time to illustrate the enabling power of the ORE specifications in the realm of scholarly citation. In the experiment, illustrated in the movie at [38], the traditional citation process is transformed to become a matter of hyperlinking a section of a digital manuscript with the URL of the to-be-cited scholarly artifact. This is similar to the manner in which non-scholarly document is hyperlinked. However, in the experiment, the authoring tool is made ORE-aware: for each URL that the author inserts into the manuscript, the tool tries to find an associated Resource Map using the Discovery techniques described in Section 3.3.2. If such a Resource Map describing an Aggregation associated with the to-be-cited artifact is found, it is scanned in search of an Aggregated Resource that provides a bibliographic description of the artifact as a representation. This is achieved by introspecting on the semantic properties of the Aggregated Resources in search of a resource with an `rdf:type` of `info:eu-repo/semantics/DescriptiveMetadata`. If such an Aggregated Resource is found, the tool de-references its URI to obtain the bibliographic description (MARC/XML in case of the experiment), parses it, and automatically inserts the citation in the References section at the end of the manuscript. In essence, the experiment demonstrates the feasibility of a new paradigm for the citation process in which no reference manager tools are used, and in which the Web itself is the reference manager database. This is an appealing illustration of one fulfillment of the object re-use goal of the ORE effort.

Once the December 2007 milestone for the release of alpha version of the ORE specifications was set, the coordinators of the ORE effort engaged with the Andrew W. Mellon Foundation in the U.S.A. and with the Joint Information Systems Committee (JISC) in the U.K. to secure funding for a limited number of small-scale experiments that have the implementation of the ORE specifications at their core, and that should result in demonstrable showcases that illustrate the enabling nature of the specifications in the realm of scholarly communication, research, and education. The Mellon Foundation funded two such projects.

- The first one, led by Michael Nelson at Old Dominion University, explores how the ORE framework can be leveraged to provide new digital preservation functionality outside of the typical repository environment. More particularly, it investigates how Resource Maps for arbitrary Aggregations can be combined with JavaScript, Wikis and email to provide a preservation function that puts client applications, such as browsers, instead of servers in the driver seat.

- The second Mellon-funded project is led by Tim Cole at the University of Illinois at Urbana Champaign. It addresses the challenge of text-on-text annotation of digitized books. Current schemes for identifying and describing annotation targets tend to be representation-specific and are expressed in idiosyncratic ways. The project investigates whether Resource Maps can be used to reveal richer targets for annotation in an interoperable and transparent way.

At the time of writing, the JISC call for proposals for ORE experiments is still open, but the outlines of one proposed project are known. The project led by Robert Sanderson and Richard Jones at the University of Liverpool and the Bristol HP Labs, respectively, will work with JSTOR to automatically produce Resource Maps for all of JSTOR's holdings. Resource Maps will go down to the page level of articles, and will express detailed resource properties wherever possible. In a next project phase, HP Labs will explore the synergy between the ORE and SWORD [3] specifications and leverage both to ingest the JSTOR Resource Maps into a DSpace repository, taking into account the rights statements for the articles expressed in those Resource Maps.

Meanwhile, other projects are starting to look into the applicability of the ORE specifications to address some of the challenges they face.

- The JournalFire project[4] that involves faculty and graduate students from several departments at the California Institute of Technology is developing an application that will allow researchers to discuss Web-based publications in online journal clubs, and to attach additional resources to those publications such as comments, keyword tags, figures, video, etc. The project is investigating the use of Resource Maps to aggregate these resources and the publication to which they pertain into a logical whole.
- The European TELplus project[5] is examining the use of Resource Maps to transfer, in a by-reference manner, OCR-ed versions of digitized books from distributed contributing sites to a centrally operated search engine.
- The DASe project[6] at the University of Texas Austin is engaging in a collaboration with the Texas Advanced Computing Center that is working on the EnVision data visualization project[7]. Envision allows authorized users to initiate simulations and/or upload large data sets resulting from simulations and to specify criteria for the visualizations. EnVision currently lacks a solution to record and maintain a consistent trail of the variety of information entities involved in creating a specific visualization, including the source data set, the parameters used for the visualization, the resulting images, and further metadata and annotations for the images. Resource Maps are being explored as a possible solution to record this aggregation of information entities, and to make it available for ingestion into the DASe system, where it would be maintained.
- Johns Hopkins University is working on a project funded by Microsoft and the Institute of Museum and Library Services to capture and associate data with publications. The project investigates the use of ORE to provide linkages between the publication, data, and other resources used to produce or process the data. Moreover, the use of ORE Resource Maps to enable content preservation and mirroring will be explored.
- The OREChem Project funded by Microsoft and involving Cambridge, Cornell, Indiana, Penn State, and Southampton is planning to use ORE as the basis of interoperability among variety of chemistry molecule repositories, and will build innovative applications on that infrastructure.

# 5. CONCLUSION AND CHALLENGES

The OAI-ORE specifications represent the evolution of DL and repository interoperability efforts so that they are more closely integrated with the Web Architecture and best practices of the Web community at large. Although the specifications have just been released, they are informed by the technologies from and experiences with both digital libraries and Semantic Web. In the same way that SiteMaps assist services by clearly enumerating the resources available at a web site, Resource Maps unambiguously enumerate distributed Aggregated Resources, and can express their types and relationships.

A number of projects are underway that are exploring different ways that OAI-ORE can be used, with a heavy emphasis on scholarly communication. Although this represents our initial target community, we hope that OAI-ORE will prove useful outside of scholarly communication and digital library environments. By embracing popular technologies such as the Atom Syndication Format, we hope to both leverage existing tools and procedures as well as provide a general solution to the problem of describing Aggregations on the Web.

# 6. ACKNOWLEDGMENTS

OAI-ORE is supported by the Andrew W. Mellon Foundation, the Coalition for Networked Information, Microsoft, and the National Science Foundation (IIS-0430906). The authors acknowledge the contributions to the OAI-ORE effort from the ORE Technical Committee, Liaison Group and Advisory Committee. Many thanks to Lyudmila Balakireva, Ryan Chute, Stephan Dresher, and Zhiwu Xie of the Digital Library Research & Prototyping Team of the Los Alamos Laboratory for their experimental work.

---